# A Compiler Infrastructure for FPGA and ASIC Development
John Demme (jdd@cs.columbia.edu), September 2016

*This whitepaper proposes a unified framework for hardware design tools to ease the development and inter-operability of said tools. By creating a large ecosystem of hardware development tools across vendors, academia, and the open source community, we hope to significantly increase much need productivity in hardware design.*

Despite the potential advantages of FPGAs and custom ASICs over CPUs and GPUs, their popularity is currently very limited. Perhaps the largest reason for this lack of usage is their cost of development, primarily in terms of programmer and designer productivity. In the FPGA case, the amount of programmer time required to correctly implement, test, debug, and optimize a large production FPGA application makes their use prohibitively expensive for all but those with the deepest pockets. Even in ASIC design, small prototypes built through multi-project services like MOSIS can reduce manufacturing costs to less than $100k, but the design costs of these projects still remain daunting. In both cases, drastic increases in productivity are necessary to increase the popularity of these platforms.

To address the problem of productivity, researchers and EDA/FPGA companies have developed tools to assist programmers. Chief among these tools are languages and HLS compilers. The ability to design a system in SystemC, Chisel, C, or even OpenCL are a real boon to productivity. Additionally, we have seen the creation of higher-level debuggers, automatic design partitioning, and utilities for automatic synthesis of memory systems and on-chip networks. Individually, these tools are valuable; together, they would make a significant dent in the productivity problem. So if all these tools have already been developed, why aren't they more popular and commonly used?

One of the major problems with the tools developed in academic environment and (to a slightly lesser extent) in industry is interoperability – the tools often cannot work together or in environments different than the ones in which they themselves were developed.  HLS compilers (academic or industrial) often produce Verilog which is only compatible with select synthesis engines or only produces efficient code for a specific set of FPGAs or ASIC tools. A particular debugger may need to communicate with a host PC and thus relies on a specific communication method (e.g. PCIe) with a specific IP component running on a specific board. Automatic synthesis tools typically produce code in a specific language (e.g. Verilog) where a potential user is using another language (e.g. Chisel) for their design, thus the produced IP may not be directly usable. Here are some real examples:

- The LegUp HLS (Canis, et al., 2011) project from University of Toronto supports only Altera. Xilinx can be targeted with a "generic Verilog" mode, which is sub-optimal.
- The CONfigurable NETwork Creation Tool (Papamichael & Hoe, 2012) from CMU produces code in Bluespec. While CMU's servers can (and do) generate Verilog for

- users, their Bluespec license only allows non-commercial research use of the generated code.
- The LEAP platform (Fleming & Adler, 2016) (a set of tools and components or – as the authors say – an "operating system" for FPGAs) is programmed in Bluespec and is thus useful only for other Bluespec programmers. Additionally, it supports only a small number of FPGA boards.
- Altera's OpenCL (Czajkowski, et al., 2012) runtime hardware assumes control over DDR and PCIe interfaces, preventing access to those interfaces by non-OpenCL code.

The problems with these tools are not because they are inherently poor tools. They stem from the many-to-many problem: it is difficult to target many different platforms or attempt to inter-operate with an arbitrary number of other tools. The solution to this problem, therefore, is to create and adopt a standard middleware for tools to build upon.

We propose to create a compiler infrastructure for hardware design, demonstrated in the below figure. It is modeled on the LLVM (Low Level Virtual Machine) project (Lattner & Adve, 2004). By defining a common intermediate representation (IR) for software, LLVM eased many practical issues of compiler and software engineering tool implementation:

- New languages need only target the LLVM IR then hand it off to LLVM which handles much of the optimization and can compile to a number of different architectures.
- When a new architecture is created, an LLVM backend for that architecture allows many languages and tools to target the new architecture rather than having to write a new compiler from scratch. Several new architectures, in fact, have quickly supported several languages rather than merely developing a custom C compiler.
- A series of new, innovative tools which operate at the IR level for program analysis and instrumentation were developed. These include memory checkers, debuggers, and parallelism discovery tools, just to name a few.

## The Low Level *Physical* Machine (LLPM) Project

Much the way LLVM (and – to some extent – GCC before it) refashioned software compiler and tools development, an analogous infrastructure for hardware could do the same for hardware tool development. Rather than developing their own compiler front-ends and optimizations, FPGA and ASIC tool vendors could target the LLPM intermediate representation, providing just a backend. They would then automatically benefit from all the languages and tools which already use the LLPM IR. Similarly, new languages, tools, and even IP libraries could target the LLPM IR, providing simultaneous access to many platforms and making inter-operability easier. Without elaborating much on the LLPM IR, here are some further, concrete examples:

- LLPM standardizes communication protocols between modules. As a result, modules written in different HDLs (e.g. Bluespec vs. Chisel) can communicate with each other easily – the frontends need not even know about one another.
- LLPM uses strongly-typed communication. Tools which automatically synthesize debuggers or host communication bridges can use type information to provide high-level interfaces to in a language-agnostic manner. This is a significant practical improvement versus providing users raw, untyped bits for manual interpretation, decoding, and encoding as existing debuggers and communication interfaces oft do.
- LLPM uses latency insensitive communication. Tools could automatically design partition, floorplan, and then create on and off chip networks to suit. Additionally, in cases where modules operate at different frequencies, LLPM-based tools could automatically synthesize multiple clocks and the necessary clock-crossing logic.
- LLPM uses an untimed model for computation within modules. Optimizations operating on this model, therefore, are not limited to maintaining the appearance of executing operations at manually-scheduled cycles. Instead, we can write optimizations which automatically pipeline. They could unroll loops. One could even convert under-utilized hardware to software or small, slower hardware to save area at the cost of performance. Few of these optimizations are novel, but none can be automatically applied to RTL and many currently get re-written for each new HLS compiler and tuned for each FPGA microarchitecture and ASIC process.
- LLPM could be backwards compatible to support many traditional RTL modules. Existing IP need only add typed, latency-insensitive wrappers. Many hardware modules already use ad-hoc LI approaches, in fact, so they need only annotate them with type information. Existing high-level compilers (e.g. OpenCL, HLS, Bluespec, etc.) could even automatically generate LLPM-compatible RTL modules. Their modules would then be able to take advantage of the LLPM tool ecosystem at and above their module boundaries.

These are just a few of the benefits of a higher-level, unified infrastructure.

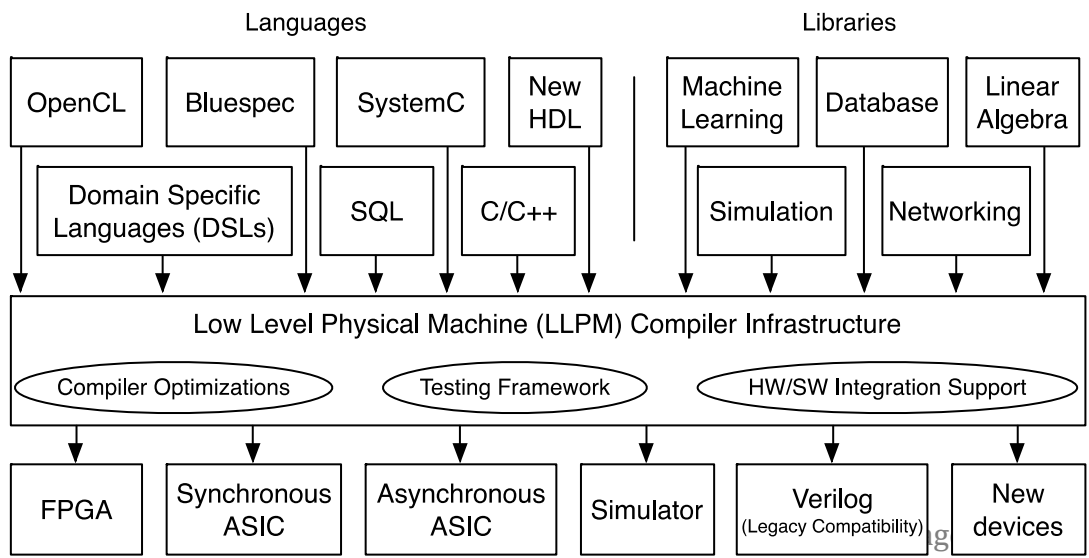

## Towards LLPM

The development of a full compiler infrastructure like LLPM would require years of development and large investment. Its full benefits would take years to be fully seen. The original paper describing LLVM, for instance, was published in 2004. Apple subsequently began investing heavily in LLVM starting in 2005. This has benefited both the community as a whole and Apple: Apple switched from GCC to LLVM starting in 2009, released their language "Swift" in 2014, and LLVM (and Clang) have become integral to Xcode, their IDE.

To develop LLPM, however, we need not replace our entire hardware design toolchain overnight to see benefits. Instead, development can be broken up into two phases: initially, a framework for working with RTL modules[1] and, later, an additional system for writing module compilers. The first phase will specify a type system, a protocol for latency insensitive communication between modules, a packaging format, and build a software infrastructure for building full systems from modules. This allows a variety of useful tools to be written without building a full compiler infrastructure. For example, the first phase could support tools like inter-module debuggers, performance monitors, host interface communication and API synthesis, multi-FPGA design partitioning, and other features which intervene and/or assist at the system rather than module granularity.

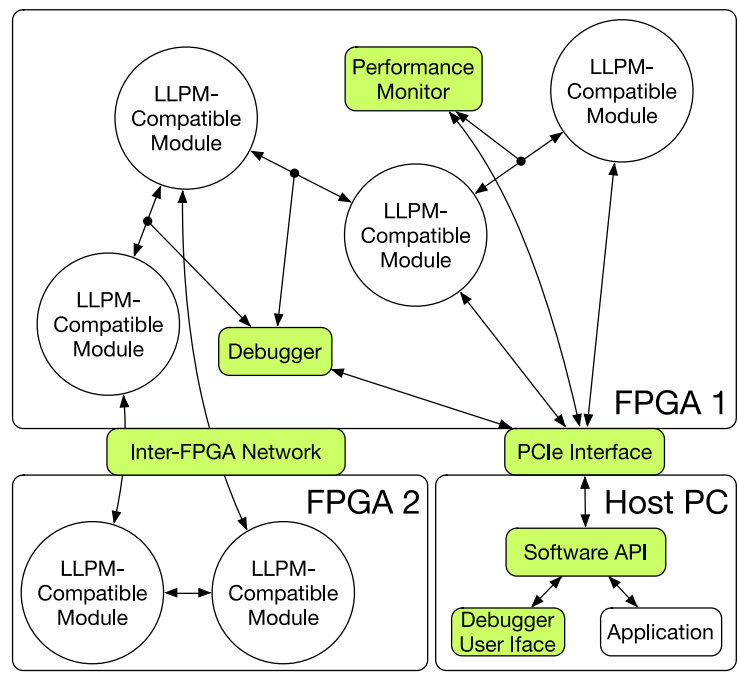

The second phase would fully develop an IR for hardware and back ends for various hardware platforms, supporting language designers and optimization researchers to fully leverage the benefits of LLPM.

***To conclude,*** it is unlikely that just one tool, optimization, or language will be the breakthrough to popularize hardware development. Rather, we posit that an ecosystem of tools is necessary. LLPM directly addresses that goal by being the foundation of this ecosystem. Thus, a project like LLPM could lead to hardware design for the masses.

---

[1] Phase 1 is very similar to MIT's LEAP project (Fleming & Adler, 2016) in many regards. In contrast to LEAP, however, LLPM is language agnostic and intended to support many languages whereas LEAP currently functions like a library for Bluespec code. Phase 1 also bears loose resemblance to QSYS and Vivado Block Designer; however, LLPM's focus is full hardware system design rather than SoC design and synthesis.